**A unified theory for the development of tinnitus and hyperacusis based on associative plasticity in the dorsal cochlear nucleus**

Holger Schulze[1] and Achim Schilling[1,2]

[1]Experimental Otolaryngology, Friedrich-Alexander University Erlangen-Nürnberg, Germany

[2]Cognitive Computational Neuroscience Group, University Erlangen-Nürnberg, Germany

Corresponding author:
Prof. Dr. Holger Schulze
Experimental Otolaryngology, Friedrich-Alexander University Erlangen-Nürnberg
Waldstraße 1, 91054 Erlangen
Germany
Holger.Schulze@uk-erlangen.de
Phone: +49-9131-8543845

**Abstract**

Tinnitus and hyperacusis can occur together or in isolation, with hyperacusis being associated with tinnitus much more frequently than vice versa. This striking correlation between tinnitus and hyperacusis prevalence implicates that there might be a common origin such as a (hidden) hearing loss and possibly interrelated neural mechanisms of pathological development of those two conditions. In this theoretical paper, we propose such interrelated pathological mechanisms, localized in the dorsal cochlear nucleus (DCN) of the brainstem, that are based on classical mechanisms of Hebbian and associative plasticity known from classical conditioning. Specifically, our model proposes that hyperacusis results from synaptic enhancement of cochlear input to the DCN, whereas chronic tinnitus results from synaptic enhancement of somatosensory input to the DCN. Specific conditions leading to one or the other condition are discussed. Our model predicts, that hearing loss leads to chronic tinnitus, while noise exposure (which may also cause hearing loss) leads to hyperacusis.

**Introduction**

Subjective tinnitus is defined as the perception of a sound in the absence of any physical sound source (Prengel et al., 2023; Kraus et al., 2016; Schilling et al., 2021). This auditory phantom percept often leads to an enormous psychic burden (Cederroth et al., 2020). In Europe the tinnitus prevalence ranges from 8,7 % to 28,3 %, with a prevalence of 12% in Germany (Biswas et al., 2022). Approximately 2% of tinnitus patients severely suffer from tinnitus and related comorbidities such as concentration issues, stress, and depression (Cederroth et al., 2020). Interestingly, 30 %-80% of tinnitus patients suffer from another pathological condition, namely hyperacusis (Pienkowski 2019). Hyperacusis refers to a reduced tolerance to sounds, which are perceived as normally loud by the majority of the population (Bigras et al., 2022). However, up to now there is no commonly accepted scientific definition (Bigras et al., 2022). In contrast to tinnitus patients where 8-28 % suffer from hyperacusis, only 5-10% of the general population suffers from hyperacusis (Pienkowski 2019). Vice versa, 86% of the hyperacusis patients suffer from tinnitus (Baguley 2003). In other words, patients may suffer exclusively from tinnitus, exclusively from hyperacusis or from both pathological conditions. The fact that both conditions can occur in isolation suggests that tinnitus and hyperacusis are caused by disjoint neural mechanisms (Schilling et al., 2023). However, the striking correlation between tinnitus and hyperacusis prevalence implicates that there might be a common origin such as a (hidden) hearing loss (Schaette and McAlpine, 2011; Norena and Chery-Croze, 2007; Knipper et al., 2013) and possibly interrelated neural mechanisms of pathological development (Schilling et al., 2023). In this theoretical paper, we propose such interrelated pathological mechanisms, localized in the brainstem, that are based on classical mechanisms of Hebbian and associative plasticity known from classical conditioning. Our model does not require any additional assumptions to explain the development of tinnitus and/or hyperacusis and therefore does not make any predictions or assumptions about possible mechanisms upstream the dorsal cochlear nucleus (DCN) like gating problems in the thalamus (Rauschecker 2024) or conscious cortical perception of a phantom sound (Sedley et al., 2016; Schilling and Krauss, 2024; Yasoda-Mohan et al., 2024). Furthermore, we will use the term hyperacusis in the sense of recruitment (Zeng 2013; Baguley 2014), which refers to the characteristic feature that mild sounds are over-amplified along the auditory pathway and perceived as too loud (Schilling et al., 2023; Kaltenbach 2007).

**Background**

There is a broad consensus that tinnitus as well as hyperacusis are induced by some kind of (hidden) hearing loss (Krauss et al., 2019; Knipper et al., 2013; Norena and Chery-Croze, 2007) Additionally, tinnitus is characterized by an increased spontaneous neural activity along the auditory pathway starting at the DCN (Kaltenbach and Afman, 2000; Kaltenbach 2007). However, tinnitus is not related to an increased stimulus evoked-activity, whereas hyperacusis in contrast is indeed characterized by increased stimulus-evoked activity (Hofmeier et al., 2021; Koops and van Dijk, 2021). Additionally, it is well known that tinnitus can occur within seconds after a short transient hearing loss (Almond et al., 2013), which is far too fast for plastic adaptations like homeostatic plasticity or so- called central gain mechanisms (Norena and Chery-Croze, 2007), which is a comparatively slow process on scales of hours or days (Zenke et al., 2017; Turrigiano 1999). In 2016, we proposed a mechanism based on a simple feedback loop that can cause the tinnitus related hyperactivity in the brainstem (The Erlangen Model of Tinnitus Development, Krauss et al., 2016; Schulze et al., 2023). The idea of that mechanism (cf. Fig. 1) is that a signal coming from the cochlea (blue) which is subthreshold e.g., due to hearing loss is lifted above the detection threshold (dashes line) by means of stochastic resonance, that is by adding neural noise which is believed to come from the somatosensory system (green) to the cochlear input. The neural noise is fine-tuned through a feedback loop that maximizes information transmission into the auditory system by maximizing the autocorrelation function of the DCN output (Krauss et al., 2016, 2017; Schulze et al., 2023) (for further explanation, refer to the figure legend of Fig. 1).

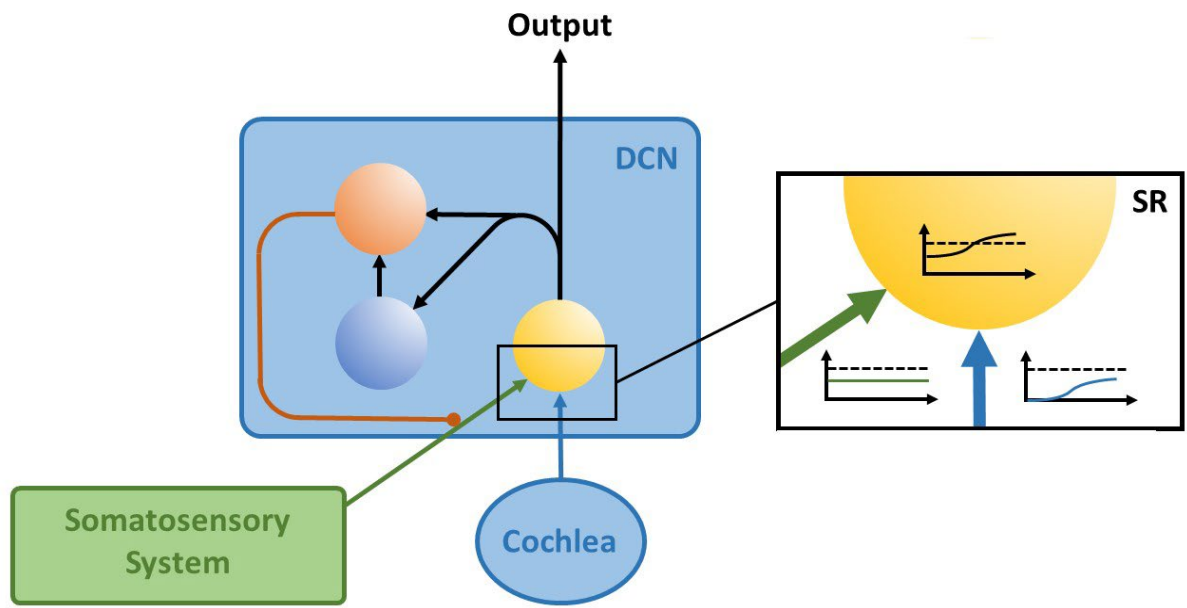

*Figure 1: The Erlangen Model of Tinnitus Development.* Like most models, ours also assumes that an initial hearing loss is the cause of the development of a tinnitus percept. The hearing loss is based on a reduction in the innervation of the inner hair cells of the cochlea (Tziridis et al., 2021), which can be so slight that no increase in the hearing thresholds can be measured in the audiogram ("hidden hearing loss", cf. Kujawa and Liberma, 2009). However, the result is still a reduced input from the cochlea into the dorsal cochlear nucleus (DCN), which can already lead to weak signals no longer triggering a supra-threshold reaction of the corresponding neurons in the DCN and thus no longer being transmitted to the subsequent auditory pathway. Our model postulates that at the level of the DCN neurons (yellow neuron) stochastic resonance (SR, enlarged section; diagrams refer to rate-intensity functions) takes place in order to enable the transmission of weak signals into the auditory pathway and thus improve hearing overall. In this process, noise in the form of spontaneous neuronal activity (green arrows) is added to the weak input signal from the cochlea (blue arrows) so that the sum of the input signal and noise is large enough to activate the DCN neuron above the threshold and thus enable the signal to be transmitted (enlarged section). The source of the neuronal noise is most likely located in the somatosensory system (Shore and Zhou, 2006). In order for SR to optimize the transmission of information into the auditory system, the amplitude of the noise (green) must be selected so that the information content of the DCN output is maximized. If this amplitude is too low, the sum of cochlear and somatosensory input does not reach the threshold of the DCN neurons; if it is too high, the signal is masked by the noise. According to our model, to determine this optimal noise amplitude, the DCN calculates the autocorrelation of the DCN output, which is a direct measure of its information content (Krauss et al., 2017), and optimizes this information content by means of a feedback loop. For this purpose, the DCN uses so-called delay lines (Licklider 1951), which superimpose a signal on itself with a time delay and can thus recognize regularly occurring patterns at fixed time intervals (which are not to be expected in noise). The output of the yellow neuron is fed directly to a coincidence detector (red neuron) and with a frequency-specific time delay corresponding to the time interval to be detected (via the blue delay neuron). The yellow neuron only reacts if it simultaneously receives input from the direct and the time-delayed projection, i.e., exactly when two action potentials occur with the desired time interval, which corresponds to the characteristic frequency in the respective frequency channel. For a characteristic frequency of 1 kHz, for example, this interval would be one millisecond. Finally, the coincidence detector in turn inhibits the somatosensory input. The greater the information content of the DCN output, the less noise from the somatosensory system has to be added to the cochlear input to ensure optimal information transmission. In this model, the perception of tinnitus always arises when the added noise from the somatosensory system itself is already supra-threshold. This means that the noise added to the cochlear input for SR from the somatosensory system not only serves to optimize the transmission of information into the auditory system, it is also the source of the neuronal hyperactivity underlying the tinnitus. Figure adopted from Schulze et al. (2023).

In classical conditioning, the association between and unconditioned stimulus (US) that is able to induce a naturally occurring unconditioned response (UR), and a conditioned stimulus (CS) is learned.

In Pavlov's original work the salivation of a dog is the UR that follows the presentation of food (US). The ringing of a bell (CS) is able to induce the same response if the CS is repeatedly presented prior to the US (Bi and Poo, 2001). This response to the bell is therefore called conditioned response (CR). Here we now incorporate the mechanistic principles known from classical conditioning into our model, so that it is not only able to explain the initial development of acute tinnitus perception, but also the pathological conditions of chronic tinnitus and hyperacusis, at the level of the DCN. Here we make no statements on higher brain structures and top-down mechanisms.

**A unified theory for tinnitus and hyperacusis development**

In this theoretical paper, we put forward the hypothesis that the Erlangen model for tinnitus development described in Figure 1 is also able to explain the development of hyperacusis. Our hypothesis is based on the notion that the essential connectivity pattern of two converging inputs into the DCN, namely one from the cochlea and one from the trigeminal system (cf. Fig. 2), mimics the basic neuronal circuit known from classical conditioning. The central idea of the theory is that – analogously to classical conditioning – the weights of these inputs to the DCN can be enhanced by means of synaptic plasticity triggered by certain input conditions and especially the relative timing between them. As a result, amplification of the cochlear input to the DCN would result in hyperacusis (HA, Fig. 2, upper right panel) while amplification of the somatosensory input would lead to chronic tinnitus (CT, Fig. 2, lower right panel).

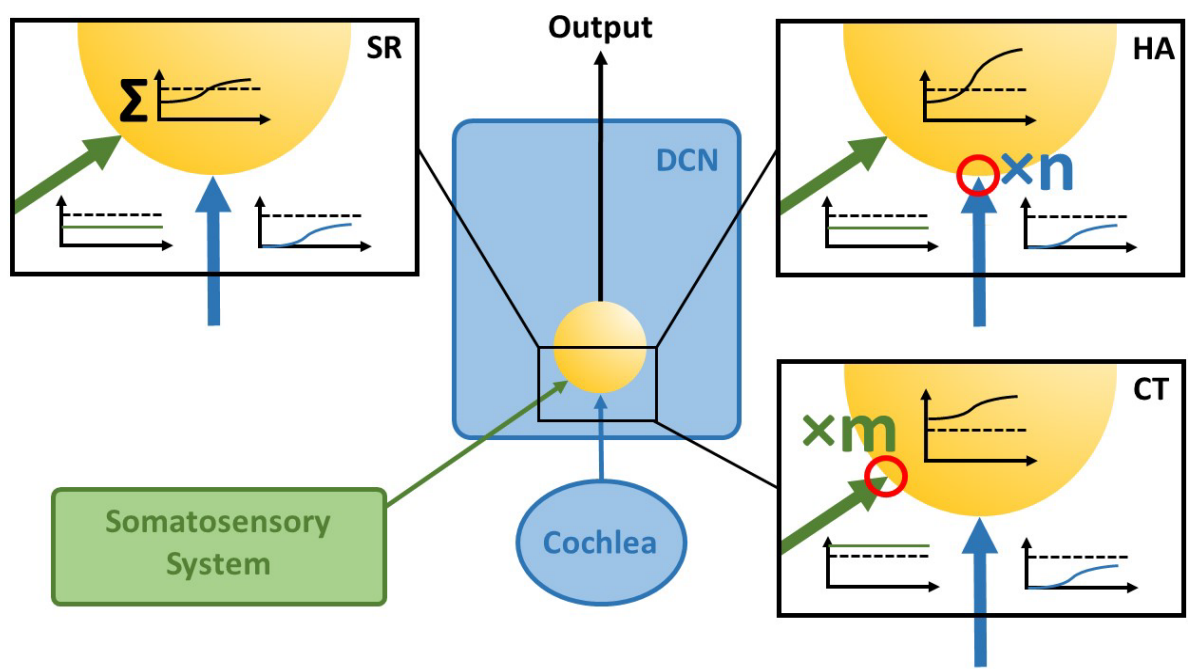

*Figure 2: Graphical abstract of the theory proposed.* The theory is based on activity dependent plastic changes of synaptic weights of inputs from the somatosensory system (i.e., the trigeminal ganglion and spinal trigeminal nucleus, Shore and Zhou, 2006) and the cochlear nerve to the DCN. Increased synaptic strength (red circles) of the cochlear input is responsible for CT, while increased synaptic strength of the somatosensory input is responsible for HA. DCN = Dorsal cochlear nucleus; SR = Stochastic resonance; HA = Hyperacusis; CT = Chronic tinnitus. For further explanations, refer to the text.

*Development of chronic tinnitus*

In our Erlangen model of tinnitus development, we have described how tinnitus-related hyperactivity develops by means of stochastic resonance (cf. Fig. 1), but we have not yet explained how the initial occurrence of such hyperactivity, i.e., acute tinnitus, can eventually lead to the condition of chronic tinnitus. In our model, acute tinnitus (Fig. 3B) occurs if the cochlear input to the DCN is reduced. This can be the case, for example, in complete silence in an anechoic chamber, where people experience transient tinnitus as long as they remain in silence. In the clinically relevant case of reduced cochlear input due to hearing loss, chronic tinnitus (Fig. 3C) may develop if the hearing loss is permanent. In this case, according to our model, the reduced cochlear input leads to a reduced information transmission into the auditory system which is detected by the DCN circuit in the form of reduced autocorrelation of the DCN output (cf. Fig. 1, 3C). The reduced autocorrelation causes a disinhibition of the somatosensory input to the DCN.

In classical conditioning, the conditioned stimulus (CS) must precede the unconditioned stimulus (US) in order to induced synaptic plasticity that results in conditioned responses (CR), i.e., strengthens the synaptic weight of the CS input (Bi and Poo, 2001). Applied to our model, this means that if cochlear input (analogue the US) that is reduced due to permanent hearing loss reaches the DCN neuron that is already activated by the disinhibited somatosensory input (analogue the CS) then the synaptic weight of the somatosensory input would be increased (Fig. 2, $x_m$; Fig. 3C, red flash) so that this input alone becomes strong enough to drive the DCN neuron alone, generating an output that is transmitted upstream into the auditory pathway and can finally be perceived as tinnitus (analogue the CR).

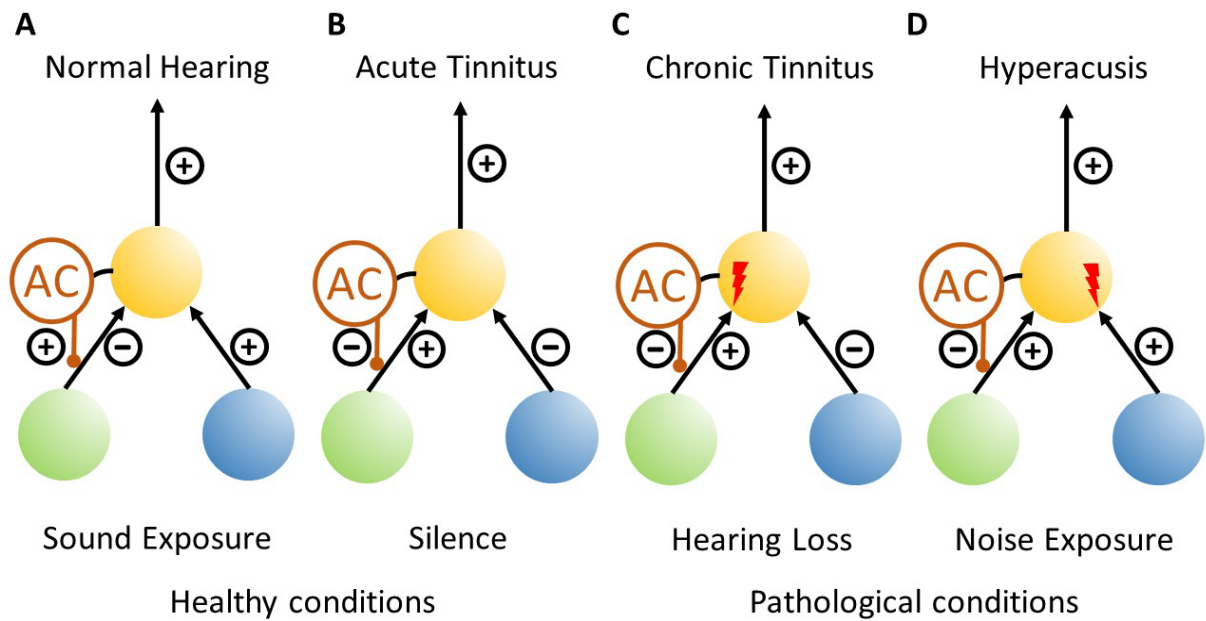

**Figure 3: The Erlangen Unified Model of Tinnitus and Hyperacusis Development.** The panels show simplified sketches of the connectivity scheme presented in Figure 1: Yellow ball: DCN input neuron. Blue ball: Cochlear input to the DCN. Green ball: Somatosensory input to the DCN. AC: Autocorrelation detector. Black lines: Excitatory connections. Red lines: Inhibitory connections. Red flash: Synaptic strengthening. For further explanation, refer to the text.

*Development of hyperacusis*

If we now take this model further, hyperacusis (Fig. 3D) can develop if the cochlear input to the DCN is amplified rather than the somatosensory input (Fig. 2, $x_n$; Fig. 3D, red flash). For this to happen, the relative timing of the two inputs is crucial. This means that if the cochlear input is to be amplified, it must precede the somatosensory input. This would occur in case of continuous noise exposure, where the cochlear input is activated, but due to the uncorrelated nature of the noise, the autocorrelation of the DCN output would still be low, which also leads to disinhibition of the somatosensory input. So, in contrast to the permanent hearing loss condition described above, where activation of somatosensory input precedes cochlear input, this relative timing of inputs to the DCN would be reversed in the case of continuous noise exposure, resulting in increased synaptic weight of cochlear input and consequently hyperacusis.

**Evaluation of the theory**

Tinnitus and hyperacusis can occur separately or together, which is easily explained by our model as it links the two pathologies to the plastic amplification of two separate synapses: according to the model, amplification of the cochlear input to the DCN leads to chronic tinnitus, while amplification of the somatosensory input to the DCN leads to hyperacusis. In cases where both inputs are amplified, tinnitus and hyperacusis occur together. In such cases, interestingly, when both conditions co-occur, increases in sound-evoked activity can be found throughout the auditory pathway (Knipper et al., 2021; Hofmeier et al., 2021). This phenomenon is also consistent with our model, as the DC shift and the greater steepness of the rate-intensity-functions (Fig. 2, insets on the right) in tinnitus and hyperacusis are superimposed here. In patients, who only have tinnitus but no hyperacusis, this superimposition and thus the additional amplification factor of the cochlear input is missing. As our model predicts, the two pathologies may reinforce each other, which is consistent with clinical data (Refat et al., 2021; Vielsmeier et al., 2020).

**Consequences of the hypothesis and conclusion**

Our model offers a new, very simple explanation for the development of both chronic tinnitus and hyperacusis and is able to describe possible interactions between the two pathologies. To our knowledge, it is the first model that describes a developmental mechanism purely in the DCN and without the involvement of higher brain areas. The fact that the plastic synaptic changes postulated here are based on mechanisms that are analogous to known learning phenomena such as classical conditioning potentially opens up the possibility of revitalizing well-known concepts such as Jastreboff's tinnitus retraining strategy (Jastreboff 2007) in order to specifically reverse the pathophysiological processes described and thus contribute to a genuine cure for tinnitus and hyperacusis.